\documentclass{epl}

\title{Crossover from Endogenous to Exogenous Activity in Open-Source Software Development}
\shorttitle{Crossover from Endogenous to Exogenous Activity...}
\author{Sergi Valverde\inst{1}}
\institute{
  \inst{1} Complex Systems Lab (ICREA-UPF), Barcelona Biomedical Research Park (PRBB-GRIB), Dr. Aiguader 88, 08003 Barcelona, Spain \\
 }


\begin{document}
\maketitle

\begin{abstract}
We have investigated the origin of fluctuations in the aggregated behaviour of an
open-source software community. In a recent series of papers [1-3], de Menezes 
and co-workers have shown how to separate internal dynamics from external fluctuations
by capturing the simultaneous activity of many system's components.  In spite of
software development being a planned activity, the analysis of fluctuations reveals how
external driving forces can be only observed at weekly and higher time scales.
Hourly and higher change frequencies mostly relate to internal maintenance activities. 
There is a crossover from endogenous to exogenous activity depending on 
the average number of file changes.  This new evidence suggests that software development is 
a non-homogeneous design activity where stronger efforts focus in a few project files.  
The crossover can be explained with a Langevin equation associated to the 
cascading process, where changes to any file trigger additional changes 
to its neighbours in the software network. In addition, analysis of fluctuations 
enables us to detect whether a software system can be decomposed in several
subsystems with different development dynamics. 
\end{abstract}

Multiple time series are available for complex systems whose dynamics is the 
outcome of a large number of agents interacting through a complex network. 
Recent measurements on the fluctuations at network nodes [1-4] indicate a 
power-law scaling between the mean$\left\langle {f_i } \right\rangle $ and 
the standard deviation $\sigma _i =\sqrt {\left\langle {\left( {f_i 
-\left\langle {f_i } \right\rangle } \right)^2} \right\rangle } $ of the 
time-dependent activity $f_i (t)$ of node $i $=1{\ldots}$N$, that is,

\begin{equation}
\sigma _i \sim \left\langle {f_i } \right\rangle ^\alpha
\label{scaling}
\end{equation}

where $\alpha $ is an exponent which can take the values between 1/2 and 1 \cite{Fluct1}. 
It seems that real systems accept a classification in two different classes 
depending on the value of this exponent. Systems with internal (or 
endogenous) dynamics like the physical Internet and electronic circuits show 
the exponent $\alpha =1/2$. On the other hand, systems either involving human 
interactions (i. e, WWW, highway traffic) or strongly influenced by external 
forces (i.e., rivers) belong to the class defined by the universal 
exponent $\alpha =1$. Interestingly, some systems display both types of 
behaviour when analysed at different scales of detail. For example, visits 
to web pages and routing of data packets in the Internet are dynamical 
processes with different origins \cite{Fluct1}. The former process is driven by user's 
demands while the latter accounts for a significant amount of internal 
activity even in the absence of human interaction (i.e., routing protocols). 

%Barabasi: "Understanding human dynamics is of major scientific and 
%practical importance and can be increasingly addressed in a quantitative 
%fashion thanks to electronic records capturing various human activity 
%patterns."

%\begin{figure}
%\onefigure[width=0.65\textwidth]{activity._eps}
%\caption{Complex fluctuations in the global activity of software development. 
%(A) Time series of global activity in the project Inkscape at the maximum time 
%resolution. The time series display intermittencies associated to sudden bursts of
%%activity. (B) The inset displays the cumulative distribution of global activity in 
%(A). Least-square fitting in log-log predicts an scaling exponent $\gamma^c = -1.64 \pm 0.02$. }
%\label{f.1}
%\end{figure}

Here, we introduce for the first time this theoretical framework to the analysis of human 
dynamics observed in open-source software development, which is an important
activity with economical and social implications. Open-source software (OSS) \cite{Raymond} 
often requires the collective efforts of a large number of experienced programmers (also 
called developers or software engineers). How individual expertise and social 
organization combines to yield a complex and reliable software system
is still largely unknown. Interestingly, many remarkable features of OSS cannot
be detected in the activity of single programmers \cite{Human}. This suggests
that, in order to understand how OSS takes place, the activities of many 
developers must be studied simultaneously.   A prerequisite to study processes of software 
change is to understand the social organization of OSS. These communities combine
two groups of people in a hierarchical or onionlike structure: (1) an inner team of 
software developers that develop and maintain the source code files and (2) the 
potentially larger community of software users (see fig.1A). This group of users triggers 
new development activities by issuing modification requests.  In addition, every software 
change has a non-zero probability to inject new software defects, which in turn may 
trigger a cascade of repair changes \cite{Challet}.

We look at software development as a sequence of software change events. Previous 
studies on software maintenance dynamics proposed a classification of changes 
in categories associated to different project stages \cite{Burch}.  These studies reported the 
frequency of every type of change. However, the software database analyzed here (see
below) does not indicate if a change addresses a user request or not. Instead, we 
suggest how the analysis of fluctuations can be used to obtain this information. 
We propose a new classification of software change as endogenous or exogenous 
depending on whether the change is independent of previous events or not.  
Because changes requested by users are independent from each other \cite{Burch}, we 
will refer to them as "exogenous".  On the other hand, cascades of correlated changes 
are "endogenous" (see fig.1A).  In a related paper, Sornette and co-authors make a distinction 
between endogenous and exogenous events in the context of book sales \cite{Sornette}.  
It was shown that exogenous and endogenous sales peaks have different relaxation 
dynamics.

\section{Data}
Detailed activity registers of the OSS community reside in centralized 
source code repositories, like the Concurrent Version System (CVS) \cite{CVS}. 
During the process of software change, developers access files to add, change 
or remove one or more lines of source code.  The CVS database tracks each file 
revision submitted by a developer.  The activity of many developers progresses in 
parallel with simultaneous changes to many files. However, the CVS system provides 
some mechanisms to ensure that any given file cannot be changed by more than 
one developer simultaneously. In addition, the CVS stores all source code files required 
to build the software system. We have shown this set of project files describes a 
complex network with an asymmetric scale-free architecture \cite{Lognets}.  
Following \cite{Lognets}, we can reconstruct this software network $G=(V,E)$ from 
the collection of source code files, where each node $v_i \in V$ represents a single 
source file and the link $(v_i ,v_j )\in E$ indicates 
a compile-time dependency between files $v_i $ and $v_j $  (see fig.1A). It can be shown that 
the number of links $L(t)$ growths logarithmically with the number of nodes $N(t)$ in 
the software network \cite{Lognets}.   Our analysis combines structural information 
provided by the software network with the time series of file changes stored in 
the CVS. We have validated our results with several software projects \cite{web}.

\section{Analysis of fluctuations}
\begin{figure}
\onefigure[width=1.0\textwidth]{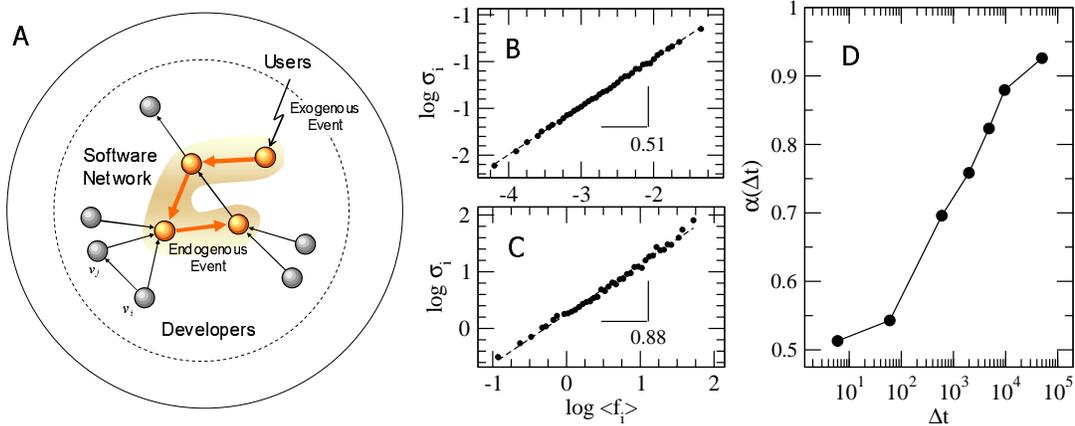}
\caption{ (A) Schematic representation of an OSS community (see text). Scaling of fluctuations 
$\sigma _i  \sim \left\langle {f_i^{\Delta t} } \right\rangle ^\alpha$ with average change 
activity for the software project XFree86, measured at different time resolutions: 
(B) $\Delta_t$=6 hours and (C)  $\Delta_t$=9600 hours. (D) shows the dependence of the 
exponent $\alpha$ with the time window $\Delta_t$. Here, the exponent $\alpha$ grows 
from 0.51 to 0.92. } 
\label{f.2}
\end{figure}

We have analysed the aggregated activity of software developers at different timescales.
Given a fixed measurement time window $\Delta t$, we measure development activity 
by looking at the dynamics of single file changes: 

\begin{equation}
f_i^{\Delta t} (t)=\sum\limits_{\tau \in \left[ {t,t+\Delta t} \right]} {c_i 
(\tau )} 
\label{eq.2}
\end{equation}

where $c_i (t)=1$ when file $v_i $ has been changed at time $t$ and $c_i (t)=0$ 
otherwise. Notice how eq.~(\ref{eq.2}) corresponds to the coarse-graining of 
the time series of file change events. In the following, we will omit the 
subscript $\Delta t$ whenever the timescale is implicit. We also define 
global activity $F^{\Delta t}(t)$ or the number of project changes at time $t$:

\begin{equation}
F^{\Delta t}(t)=\sum\limits_{i=1}^N {f_i^{\Delta t} (t)} .
\end{equation}

In figure \ref{f.2}D we display the scaling of fluctuations with the average activity 
(see eq.~(\ref{scaling})) in a software project at different time scales. There is a 
dependence of the scaling exponent with the time window $\Delta t$. 
The observed exponent is less than 1 for a wide range of time scales (see \ref{f.2}B) ,
thus suggesting and endogenous origin of development activity.  On the other hand, the analysis of
fluctuations in various OSS projects at monthly and large time scales yields an 
exponent closer to 1 (see \ref{f.2}C). The external driving force becomes stronger when $\Delta t$ 
increases.  In the following, we further investigate the origin of fluctuations in 
software development dynamics with a more robust measure.

\section{Crossover in internal dynamics}

We can determine if OSS dynamics has an endogenous or exogenous origin by 
separating internal and external contributions \cite{Fluct2}.  We split the timeseries of 
individual file changes $f_i (t)$ in two different components: (i) internal fluctuations $f^{int}(t)$ 
governed by local interaction rules and (ii) external fluctuations $f^{ext}(t)$ caused by 
environmental variations, that is, 

\begin{equation}
f_i (t)=f^{int}(t)+f^{ext}(t)
\label{eq4}
\end{equation}

where the external activity $f^{ext}(t)$ represents the expected fraction of changes
shared by file $v_i $:

\begin{equation}
f_{^i }^{ext} (t) = A_i \sum\limits_{i = 1}^N {f_i (t)} 
\label{eq5}
\end{equation}

\begin{figure}
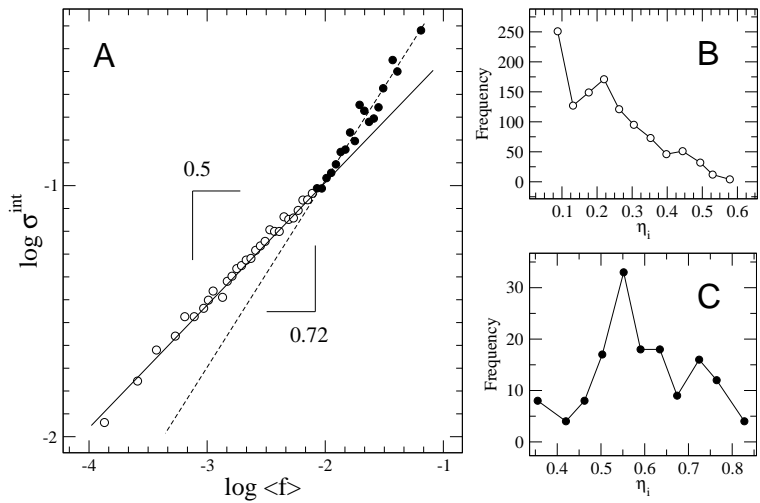

\onefigure[width=0.7\textwidth]{fig2.eps}
\caption{  (A) Crossover observed in the scaling of internal fluctuations with 
average flux, around $10^{-2}$, for the Apache project. In (B) and (C) we show the binned distribution of ratios for 
the two project file subsets $\left\langle {f} \right\rangle < 10^{-2}$ (open circles) and 
$\left\langle {f} \right\rangle > 10^{-2}$ (black circles), respectively. In all plots, 
$\Delta t$ = 10 hours.} 
\label{f.3}
\end{figure}

Here $A_i$ is file centrality  \cite{Fluct2}, defined as the overall fraction of 
changes received by the file $v_i$:

\begin{equation}
A_i  = \frac{{\sum\nolimits_{t = 1}^T {f_i (t)} }}{{\sum\nolimits_{t = 1}^T {\sum\nolimits_{i = 1}^N {f_i (t)} } }}
\label{centrality}
\end{equation}

and $T$ is the timespan of software development. Notice that file centrality $A_i$
is independent of the observation window $\Delta t$. By definition, external 
fluctuations allways scale linearly with the average number of file changes, 
$\sigma^{ext} \sim \left\langle {f} \right\rangle$. On the other hand, the 
exponent $\alpha$ governing the scaling of internal fluctuations with average 
flux $\sigma^{int}  \sim \left\langle {f} \right\rangle ^{\alpha}$ indicates if dynamics 
has an endogenous ($\alpha = 0.5$) or exogenous ($\alpha= 1$) origin. 
Interestingly, we observe a crossover in the internal activity of open-source 
software development depending on the average number of file changes 
$\left\langle {f} \right\rangle$ (see fig. \ref{f.3}A).  The crossover is less visible at
large time scales $\Delta t$.

The analysis of single node fluctuations provides additional evidence for this crossover.
The ratio $\eta _i =\sigma _i^{ext} /\sigma _i^{int} $ between external and internal fluctuations 
indicates wether node dynamics is external ($\eta _i >>1)$ or internal ($\eta _i <<1)$. In 
order to characterize the system's overall behavior, we can compute the 
distribution of ratios $P(\eta _i )$ \cite{Fluct2}. This measure was 
shown to be robust to variations in the measurement time window $\Delta t$.  For example, 
figure \ref{f.3} displays the distribution of ratios $P(\eta _i )$ measured in two
different subsets of files in the Apache project. On the one hand,
we can see that $P(\eta_i)$ is peaked around $0.55$ (see fig. \ref{f.3}C) for the subset of files 
with $\left\langle {f} \right\rangle > 10^{-2}$. This suggests exogenous activity in 
 a core set of project files (those depicted with black circles in fig.\ref{f.3}A and fig.\ref{f.3}C). 
Moreover, $P(\eta_i)$ is skewed towards lower ratios (around $0.1$) for project files 
with $\left\langle {f} \right\rangle < 10^{-2}$ (white circles in fig.\ref{f.3}A and fig.\ref{f.3}B) . 
On the other hand, activities involving less changed files have an endogenous origin (see fig. \ref{f.3}B).

\section{Propagation of Changes}

\begin{figure}
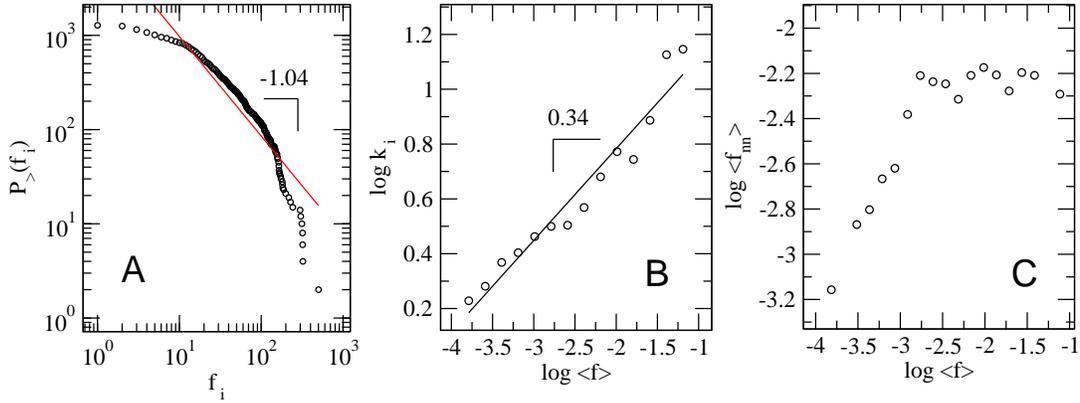

\onefigure[width=1.0\textwidth]{apache_centralities2.eps}
\caption{ Measuring internal propagation of changes in the Apache project. (A) The cumulative total 
activity distribution $P_>(f_i)$ is broad scale. (B) Scaling of average activity/flux with node degree, 
$k_i \sim \left\langle f \right\rangle ^{\beta}$, with $\beta \approx 0.34$.
(C) Average neighbors activity $\left\langle f_{nn} \right\rangle$ scales with average node activity 
$\left\langle f \right\rangle < f_0$ and then saturates $\left\langle f_{nn} \right\rangle=const$ 
once the crossover $f_0 = 10^{-2.5}$ is reached, $\left\langle f \right\rangle > f_0$.
In order to reduce the noise, data have been logarithmically binned in (B) and (C) plots. The measurement 
window is $\Delta t$ = 10 hours.} 
\label{f.4}
\end{figure}

Crossover in internal fluctuations stems from the inhomogeneous nature
of software development. A large development effort aims to a small number of
core files, which change more frequently than other project files.
In a related paper, network heterogeneity was shown to have an impact in
the dynamics of diffusion processes \cite{Yook2005}. When the 
diffusive process is multiplicative and the underlying topology is intrinsically 
inhomogeneous, there is a crossover from $\alpha = 0.5$ to $\alpha = 1$ in
the scaling of fluctuations with the average flux (eq.(\ref{scaling})). 
Such diffusive network processes can be modeled through the Langevin equation by a 
mean-field approximation \cite{Yook2005}. The change of mass at node $i$
during a unit time interval is:

\begin{equation}
f_i (t + 1) = f_i (t) + \sum\limits_j^{k_i } {\frac{1}{{k_j }}\eta _j (t)f_j (t)} 
\label{Langevin}
\end{equation}

where the second term represents the incoming mass from the nearest
neighbors and $\eta_j(t)$ is a uniform random variable (i.e., multiplicative noise
term). Because we are focusing in the internal diffusion process we do not take 
into account additional terms like outgoing mass and/or uncorrelated 
Gaussian noise.  This type of diffusion processes display a characteristic 
scaling in the probability distribution $P(f_i) \sim  {f_i}^{-1-\mu}$ \cite{Yook2005}.
The continuous approximation of the previous equation is

\begin{equation}
\frac{{df}}{{dt}} \cong \sum\limits_j^{k_i } {\frac{1}{{k_j }}\eta _j (t)f_j (t)} 
\cong \left\langle {k_i } \right\rangle \frac{1}{{\left\langle {k_{nn} } \right\rangle }}\left\langle {\eta _j (t)f_j (t)} \right\rangle 
\end{equation}

where $\left\langle {k_{nn} } \right\rangle$ denotes the average degree
of a node's nearest neighbors. Because $\eta_j(t)$ and $f_j(t)$ are independent
variables and assuming that $\left\langle {k_{nn} } \right\rangle$ is function of $\left\langle k \right\rangle$:

\begin{equation}
\frac{{df}}{{dt}} \cong \frac{{\left\langle k \right\rangle }}{{\left\langle {k_{nn} } \right\rangle }}\left\langle {\eta _j } \right\rangle \left\langle {f_{nn} } \right\rangle  \equiv J\left( {\left\langle k \right\rangle } \right)\left\langle {f_{nn} } \right\rangle 
\label{MeanField}
\end{equation}

 where $\left\langle {f_{nn} } \right\rangle$ indicates the average incoming mass in
the nearest neighbors of a node.  For the Barab\'asi-Albert network \cite{Barabasi}, the numerical solution 
of the above equation shows that $\left\langle f_{nn} \right\rangle $ 
decreases as $\left\langle  f \right\rangle$ increases and then saturates to a constant value 
for $\left\langle  f \right\rangle  > f_0$ (see \cite{Yook2005} for details). The observed value of 
$f_0$ indicates the crossover between endogenous and exogenous dynamics.

Interestingly, model requeriments (i.e, diffusion process on a heterogeneous network) are 
met by software projects.  Empirical studies of software
maintenance reported that change propagation is a central feature of software maintenance 
\cite{change}. Propagation is necessary because there are dependencies between project files and
developers must ensure that related files are changed in a consistent way. Recall the
software network $G$ captures these file dependencies (see above). The software network 
displays a scale-free structure due to extensive reuse during software development \cite{Lognets}. 

Furthermore, our measurements on real OSS projects seem consistent with model predictions.
We have observed that, for all software projects analyzed here, the propability distribution 
$P_>(f_i)$ has a long tail. For example, power-law fitting for the Apache project predicts an 
exponent $-1-\mu \approx -2.04$ for the incoming flux distribution (see fig. \ref{f.4}A, cumulative 
probability distribution is used to reduce the impact of noisy fluctuations).  As hypothesized above, 
the plot in fig. \ref{f.4}B shows that key files having a large number 
of dependencies are changed more frequently. We have checked that $\left\langle {k_{nn} } \right\rangle$ is 
a function of $\left\langle k \right\rangle$ (not shown here). As seen in fig. \ref{f.4}C, the average 
neighbour activity increases with average node activity $\left\langle f \right\rangle$ and it is approximately constant 
$\left\langle {f_{nn} } \right\rangle \approx const$ for $\left\langle f \right\rangle > f_0$ with $f_0=3.16 * 10^{-3}$.
This value of $f_0$ is consistent with the observation made in fig. \ref{f.3}A. In this case, eq. (\ref{MeanField}) 
predicts $\left\langle f \right\rangle \sim \left\langle k \right\rangle ^ {\alpha} $ with $\alpha = 1$ to be 
compared with the measured exponent 0.72 (see fig. \ref{f.3}A).

\section{Different subsystems display different scaling laws}

\begin{figure}
\onefigure[width=0.95\textwidth]{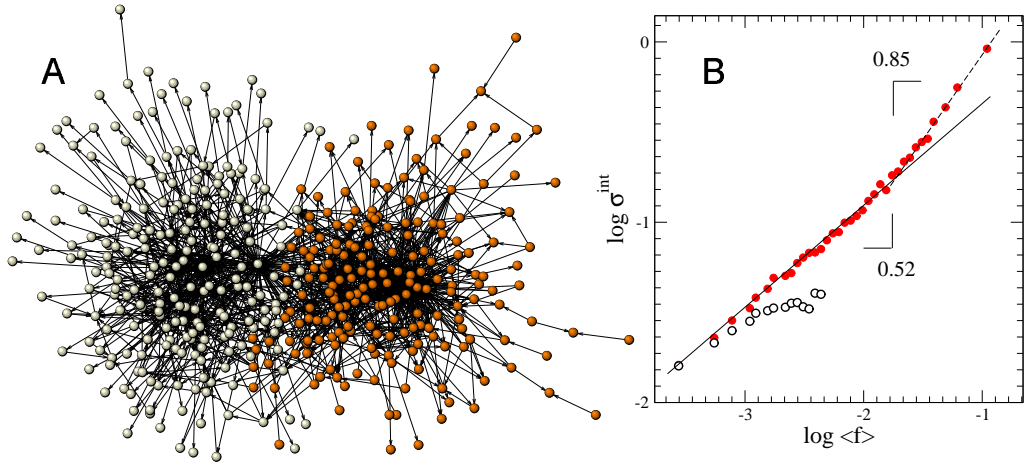}
\caption{Scaling of internal fluctuations in different subsystems of the TortoiseCVS software project. (A)
Modular organization of the corresponding software network, where node represents files and
links depict dependencies. Nodes within the same subsystem are displayed in the same colour. 
(B) Different scaling laws of internal subsystem fluctuations with average flux, 
$\sigma_i \sim \left \langle f_i \right \rangle ^{\alpha_{int}}$, for the main application 
subsystem (black balls) and for the window subsystem (so-called {\em wxwin}, white balls).  }
\label{tortoisenet}
\end{figure}

A practical application of fluctuation analysis is the identification of files that change
together \cite{Gall}. This suggests a method for community detection based 
on individual node dynamics. In our context, we have observed that some subsystems are 
characterized by different scaling laws in their internal fluctuations with average activity.
For example, figure \ref{tortoisenet} summarizes the analysis of internal fluctuations in 
the software project TortoiseCVS.  There are two clearly defined subsystems,
the main application subsystem (dark balls) and the window library {\em wxwin} 
(white balls), characterized by different change dynamics (see fig. \ref{tortoisenet}A).
The crossover behaviour can be appreciated in the scaling of internal fluctuations for the main 
TortoiseCVS subsystem (the exponent for $\left\langle f \right\rangle > f_0$  is 
$\alpha_{int} \approx 0.85$, see fig. \ref{tortoisenet}B). 
The main subsystem concentrates the largest fraction of changes. On the other hand,
the crossover is not observed in the scaling for the {\em wxwin} subsystem 
(see fig. \ref{tortoisenet}B), which is an utility library imported from an external development
team. The minimal amount of activity regarding the {\em wxwin} subsystem 
(sporadic changes in the library communicated by the external team and minor adjustments
required by the main subsystem) suggests an explanation for the absence of a crossover. 

In short, we have provided empirical and theoretical evidence for a well defined crossover 
in the dynamics of software change. This is the first reported example of such behaviour in a large-scale
technological system. It shows that OSS systems exhibit some traits in common with
other complex networks. The presence of crossover allows to distinguish between 
internal and external components of the dynamics and then provides a powerful 
approach to uncover the relative importance of exogenous versus endogeneous
dynamics.
\section{Conclusions}

\acknowledgments
Sergi Valverde dedicates this paper to his daugther Violeta. We thank Ricard Sol\'e and Damien Challet.
This work has been supported by the EU within the 6th Framework Program under contract 001907 (DELIS).

\end{document}